\documentclass[showpacs,showkeys,aps,amsmath,amssymb]{revtex4}

\usepackage[final]{graphicx}
\usepackage{t1enc}
\usepackage{bm}
\usepackage{amsmath,amssymb}

\newcommand{\beq}{\begin{equation}}
\newcommand{\eeq}{\end{equation}}
\newcommand{\beqa}{\begin{eqnarray}}
\newcommand{\eeqa}{\end{eqnarray}}

\begin{document}

\title{\bf The topological hypothesis on phase transitions: 
the simplest case}

\author{Ana C. Ribeiro Teixeira}
\email{anacarol@if.ufrgs.br}

\author{Daniel A. Stariolo}
\altaffiliation{Research Associate of the Abdus Salam International
Centre for Theoretical Physics, Trieste, Italy}
\email{stariolo@if.ufrgs.br}
\affiliation{Departamento de F\'{\i}sica\\ 
Universidade Federal do Rio Grande do Sul\\ 
CP 15051, 91501--979, Porto Alegre, Brazil}

\date{\today}

\begin{abstract}
We critically analyze the possibility of finding signatures of a phase transition 
by looking exclusively at static quantities of statistical systems, like e.g., the topology of 
potential energy sub-manifolds (PES). This topological hypothesis has been successfully tested in a
few statistical models but up to now there is no rigorous proof of its general validity.
We make a new test of it analyzing the, probably, simplest example of a
non trivial system undergoing a continuous phase transition: the completely connected
version of the spherical model. Going through the topological properties of its PES it is shown 
that, as expected, the phase transition is correlated with a change in their topology. 
Nevertheless this change, as reflected in the behavior of a particular topological invariant, 
the Euler characteristic, is small at variance with the strong singularity observed in 
other systems. 
Furthermore, it is shown that in the presence of an external field, when the phase transition
is destroyed, a similar topology change in the sub-manifolds is still observed at the maximum value 
of the potential energy manifold, a level which nevertheless is thermodynamically inaccessible.
This suggests that static properties of the PES are not enough in order to decide whether a phase
transition will take place, some input from dynamics seems necessary. 
\end{abstract}


\maketitle

\section{Introduction}

In a series of interesting papers \cite{CaiCaClPe97,FrCaSpPe99,FrPeSp00,CaPeCo03,AnCaPeRuZa03} 
appeared 
in the last few years the possibility has been advanced that phase transitions may be signaled by 
suitable changes in some topological properties of the configuration space manifold. This implies a
 different approach to
phase transitions from the classical one based on singularities of thermodynamic potentials. 
The topological hypothesis implies a static point of view on phase transitions for it is based only
 on properties of static quantities, like the potential energy manifold. Under
rather general conditions dynamics seems to play no role \cite{CaPeCo00}. This is clearly a strong
assumption and, if confirmed at least for a restricted class of systems, should provide
a new and powerful approach to understand the microscopic mechanisms underlying a phase transition.
Up to now the topological hypothesis has been verified in a few models, most notably the
Hamiltonian XY mean field model \cite{CaCoPe02,CaPeCo03}, the two dimensional lattice
$\phi^4$ model \cite{FrPeSp00}, the $k$-trigonometric model \cite{AnCaPeRuZa03} and recently in 
the Bishop-Peyrard model of DNA denaturation \cite{GrMo04}. The topology of
 these models was investigated by calculating a topological invariant, the Euler
characteristic $\chi(v)$ defined on sub-manifolds $M_v$ of the potential energy manifold :
 $M_v \equiv \{q \in \mathcal{R}^N|V(q) \leq v \}$, 
where $V(q_1,\ldots ,q_N)$ is the potential energy function of the system. In all the
previous models the Euler characteristic shows a strong singularity at a critical level
$v_c$ in correspondence with the critical values of the energy and temperature
at the phase transition $e_c = T_c/2\, + <v(T_c)>$ ($<v(T_c)>=v_c$). In the Hamiltonian XY
mean field model and in the $k$-trigonometric model, which is also a mean field one, 
it was observed that $\lim_{N \to \infty} \frac{1}{N} \log{|\chi_v|}$ is singular at $v_c$. 
After these evidences the question that remains to be answered refers
to the necessary and sufficient conditions in order that a topology change of sub-manifolds
of the configuration space reflects the presence of a phase transition. Recently a theorem
was proved stating that a topology change of configuration space is in fact necessary 
\cite{FrPeSp03,FrPe04} for
a phase transition to occur. The theorem covers a wide class of systems with smooth, finite range 
and confining potentials bounded from below. But, although at a phase transition a
topology change must necessarily happen, the converse is not true. For it is known that topology
changes are common in configuration space while they do not necessarily imply the presence
of a phase transition. The next, more difficult task, is to find the sufficient conditions 
in order to relate topology changes with phase transitions. Up to now the only hints 
about what those conditions could be come from the observed behavior in the exactly
solved XY and $k$-trigonometric models. Specially after the results in the XY model
the authors conjectured that to entail a phase transition the topology change
must involve the attachment of handles of $\mathcal O$(N) different types on the same
critical level \cite{CaPeCo03}. 
We will show below a much simpler system undergoing a phase transition in which 
this mechanism is not present.

In this work we study the connection between topology of the potential energy manifold 
and the thermodynamics of a very simple model: the completely connected ferromagnetic spherical 
model. The potential energy manifold of this system is a hyper-sphere and its
topology is therefore trivial. We firstly discuss the model without external field. 
The critical points of the energy function within the domain of the spherical constraint are two 
isolated symmetric minima corresponding
to the ground states of the system and a highly degenerate maximum. There are no saddle
points. The sub-manifolds $M_v$ at fixed $v$ correspond to two disconnected $(N-1)$-dimensional 
disks, joining each other and completing the whole sphere at the maximum level $v_c$. This level
coincides with the critical value obtained from thermodynamics implying the coincidence
between the topology change, the closing of the hyper-sphere, and the phase transition, in
agreement with the topological hypothesis. Nevertheless we will see that the Euler 
characteristic presents at best only a small discontinuity at the transition point and also only
one handle is attached at the upper critical level which corresponds to the phase transition. 
This suggests
that the sufficiency condition discussed above does not necessarily relates with the
behavior of the Euler characteristic at the transition. 
In fact the topology change at the transition level in this model 
seems to be small in the sense that only one handle is attached in order to complete the whole
manifold. Then we discuss the model in the presence of an external field. The topology is
essentially the same as in the previous case, except that the symmetry between the two minima is 
broken. Now
one minimum is the ground state and the other, a meta-stable state. We show that a topology change
still exists where the hyper-sphere closes itself, but more importantly,
this sector of the manifold is inaccessible to the physical system. As a consequence, this topology
 change (similar to the one occurring when there is no external field) cannot be related 
with a phase transition, which in fact is destroyed by the presence of a finite field. In order
to conclude this we need to add some information from the whole problem, e.g. the
knowing of the caloric curve, which depends on the dynamics. This suggests that topology
alone may be not enough to decide whether a phase transition will take place for a given
interaction potential.

\section{The model}
We studied the completely connected version of the classical spherical model 
introduced by Berlin and Kac \cite{BeKa52}. It consists of a set of N classical
spin variables $\{s_i \in\mathcal{R}, i=1,\ldots ,N\}$ which interact through 
the potential energy function:
\beq
V = -\frac{J}{2N}\sum_{i\neq j}^N s_i\,s_j - H \sum_{i}^{N} s_i
\label{energy}
\eeq
and the spins are subject to a spherical constraint:
\beq
\sum_{i=1}^N s_i^2 = N.
\eeq
The exchange coupling $J > 0$ corresponds to a ferromagnetic interaction, $H$ 
is an external field and the factor $1/N$ in the energy function is needed in 
order to make the model extensive in the thermodynamic limit.

The thermodynamics of the model can be computed exactly following  closely the
original solution of Berlin and Kac for the finite dimensional version. 

\subsection{Zero external field}
\begin{figure}[ht]
\includegraphics[width=10cm,height=7cm]{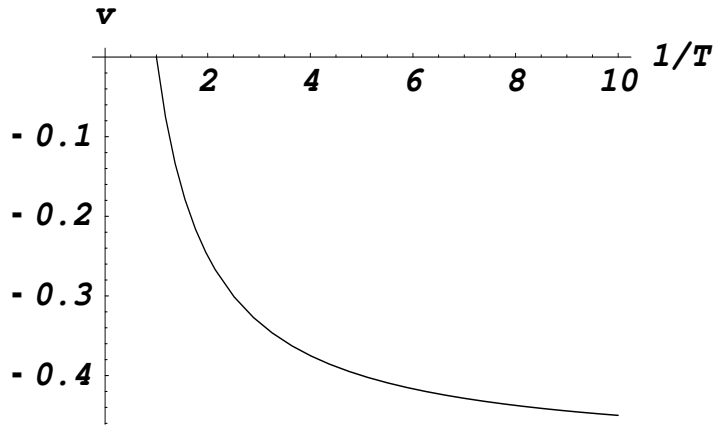}
\caption{Internal energy per particle as a function of inverse temperature in the completely connected spherical model for zero external field $(J=1)$.}
\label{fig.energia.sem.campo}
\end{figure}

For $H=0$ a saddle point approach leads to a Curie-Weiss critical point at 
$\beta_c J=1$, where $\beta=1/T$. The internal energy per particle behaves as:
\beq
v = \left\{
           \begin{array}{rl}
                        \frac{1}{2\beta} - \frac{J}{2}, & \,\,\, T < T_c\\
                        0, & \,\,\, T > T_c
            \end{array} \right. \label{mean_v}
\eeq
Consequently at the critical point the mean potential energy is $v_c\equiv v(\beta_c)=0$ as shown in figure \ref{fig.energia.sem.campo}.

This is the relevant information needed on the phase transition in this model in order to test the topological hypothesis. Below it is shown that the level $v_c$ corresponds to the maximum of the 
potential energy per particle and precisely at this level a topological change takes place in the 
sub-manifolds $M_v$. 

\subsection{Finite external field}
\label{H.dif.0}

The saddle point approach when $H \neq 0$ leads to a saddle point equation which has a finite solution for any finite temperature. Consequently the phase transition is destroyed by the field \cite{BeKa52}. In the thermodynamic limit the internal energy per particle  is given by:
\beq
v = \frac{1}{2\beta} - J\, z_s - \frac{H^2}{4J(z_s-1/2)}
\eeq
where $z_s$ is the solution of the saddle point equation:
\beq
\frac{1}{z_s} + \frac{\beta H^2}{2J(z_s-1/2)^2} - 2\beta J = 0
\eeq

\begin{figure}[ht]
\includegraphics[width=10cm,height=7cm]{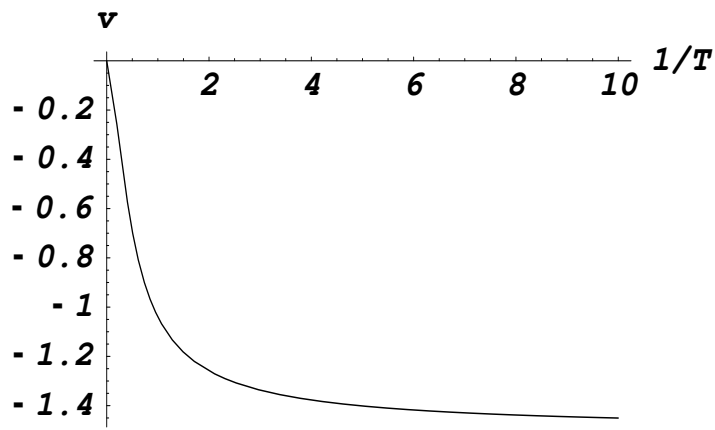}
\caption{Internal energy per particle as a function of inverse temperature for an external 
field $H=1$ $(J=1)$.}
\label{fig.energia.com.campo}
\end{figure}

A plot of the internal energy $v$ as a function of $\beta=1/T$ is shown in figure \ref{fig.energia.com.campo}. When $\beta \to \infty$, $v \to -3/2$, which is the energy of the ground state for $H=1$. For $\beta \to 0$, $v \to 0$, a value below the maximum of the potential energy per particle which is $v_{max}=H^2/2=1/2$ in this case. The conclusion is that the system is unable to reach the levels of 
potential energy above $v =0$, for arbitrarily growing temperature. This is connected with the fact that the phase transition is absent when $H$ is finite. This behavior is reflected in the topology of the accessible sub-manifolds of potential energy, as will be seen below.

\section{Critical points and topology of the potential energy manifold}
\subsection{Zero external field}
From the previous definition of the model it is clear that the whole energy manifold is a 
hyper-sphere in $N$ dimensions, or in topological language an $(N-1)$-sphere. At a given level of 
potential energy $v$ the accessible sub-manifold is represented by the intersection of the energy 
surface (\ref{energy}) with the hyper-sphere. The topology of the energy function is easily revealed
 by diagonalizing the quadratic form. We can write \cite{BeKa52}:
\beq
\sum_{i,j} s_i\,s_j = {\bf s^T\,M\,s}
\eeq
with ${\bf M}$ the symmetric matrix with all elements equal to one except for the diagonal ones 
which are zero. This matrix is symmetric and is therefore diagonalized by means of an orthogonal 
transformation ${\bf V}$ such that:
\beq
{\bf M\,{V}_k} = \lambda_k {\bf {V}_k}, \,\,\,\,\, {\bf V^TV=V^{-1}V=\mathcal{I}},
\eeq
where ${\bf {V}_k}$ is the k column of the transformation matrix.

Applying this transformation to ${\bf M}$:
\beq
{\bf s^T\,M\,s} = \sum_{i=1}^N \lambda_i\, y_i^2
\eeq
where ${\bf y=V^T s}$. The spherical constraint is invariant:
\beq
\sum_{i=1}^N s_i^2 = \sum_{i=1}^N y_i^2
\eeq
From now on we will work in the base which diagonalizes ${\bf M}$. The eigenvalues $\lambda_i$ can 
be readily computed:

\beqa
\lambda_1 & = & N-1 \nonumber\\
\lambda_k & = & -1 \,\,\,\,\,\,\,\,\, k=2,\ldots,N.
\eeqa
The matrix ${\bf M}$ possesses one single positive  eigenvalue and $N-1$ negative degenerate ones. 
In the base $\{y_i\}$ the energy function can be written:
\beq
V = -\frac{J(N-1)}{2N}\, y_1^2 + \frac{J}{2N}\sum_{i \geq 2} y_i^2
\label{ene_diag}
\eeq
For simplicity in what follows we fix $J=1$. Now the energy per particle is limited between 
$1/2N-1/2 \leq v \leq 1/2N$, or in the thermodynamic limit $-1/2 \leq v \leq 0$. In order to get the
 critical points of V on the $(N-1)$-sphere we introduce a Lagrange multiplier to enforce the 
spherical constraint and define:
\beq
F = V + \mu(\sum_{i=1}^N y_i^2 - N)
\label{function_F}
\eeq
Now the critical points are given by $\frac{\partial F}{\partial y_i}=0$, which give:

\beqa
 \left( 2\mu-\frac{N-1}{N}\right) y_1 & = & 0 \nonumber \\
 \left( 2\mu + \frac{1}{N} \right) y_i & = & 0, \,\,\,\,i\neq 1
\label{cript_F}
\eeqa
From this we have two possibilities, either $\mu=(N-1)/2N$, which gives two isolated minima
$\{y_1=\pm \sqrt{N},\,\,y_i=0,\,\,i\neq 1\}$, or $\mu=-1/2N$, which gives
$\{y_1=0,\,\, \sum_{i\neq 1}y_i^2=N\}$ corresponding to a degenerate maximum, completing 
the $(N-1)$-sphere.
Consequently, the potential energy manifold has only two critical sub-manifolds corresponding to the
 minimum and maximum values of the function. There are no saddle points. This structure is trivial 
and allows us to visualize immediately the topology changes as the level $v$ is increased. This 
triviality is a property of the completely connected model only. In finite space dimensionality the 
potential energy manifold is more complex and already for $d=1$ it shows a non-trivial structure of 
saddle points \cite{ana2}.

The natural framework for analyzing the relation between critical points and topology changes in a 
manifold is Morse theory \cite{NaSe}. Because of the simplicity of the spherical model one can make 
a very intuitive analysis of the topology changes in this case without resorting to Morse theory. 
The topology of the model is analyzed in the context of Morse theory in the Appendix.

Although we are interested in the behavior of the system for high dimensionality $N$, there is only 
one direction, namely $y_1$, which breaks the spherical symmetry of the potential energy function 
and the problem can then be effectively analyzed in a two dimensional plane spanned by $y_1$ and
any other orthogonal direction. Without loss of generality, we will consider directly the case
with $N=2$.

\begin{figure}[ht]
\includegraphics[width=10cm,height=8cm]{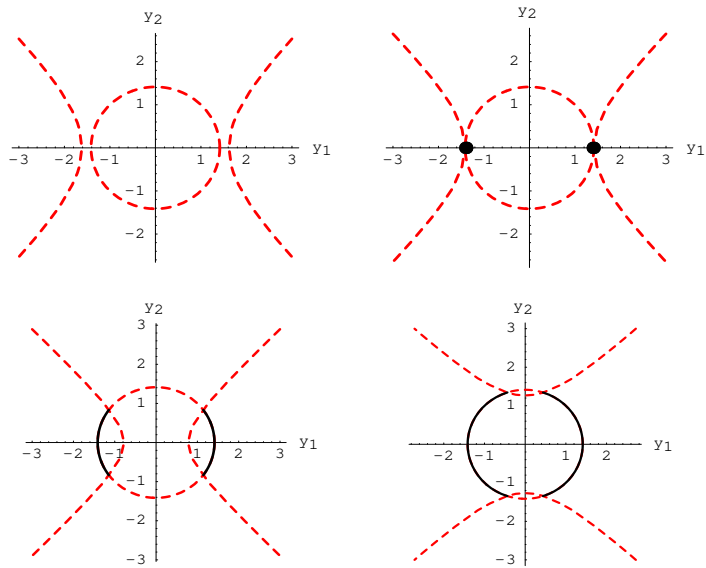}
\caption{(Color online) 
Evolution of the potential energy manifold of the $N=2$ spherical model for four levels 
$v=V/N$. The colored dashed lines represent the spherical constraint (a circle for N=2) and a 
particular level set of the function $v$. The real sub-manifolds depend on the level $v$ and are 
the continuous black sectors. Top left: $v=-0.325$, the manifold is empty. Top 
right: $v=-0.25$, the manifold emerges at the two black dots (ground states). Bottom left:
$v=-0.075$, two symmetric sectors of the sphere are accessible. Bottom right: $v=0.2$, near the
transition the manifold is nearly completed and a large fraction of the sphere is accessible.}
\label{fig.manifold}
\end{figure}

In figure (\ref{fig.manifold}) it is shown the evolution of the sub-manifolds $M_v$ for four 
increasing values of $v$. The top left panel corresponds to a level $v$ below the minimum of the 
potential energy per particle $v<-0.25$. 
In this case the manifold is empty, this is a forbidden region for the system. As the system crosses
 the level $v=-0.25$ a first topology change happens (top right panel). At this level two points are
 accessible in configuration space, the symmetric ground states of the system. Above this level the 
sub-manifold $M_v$ is diffeomorphic to two disconnected (hyper)-disks
(in high dimensions). This situation is represented by the bottom left panel in figure 
(\ref{fig.manifold}). Note that in the $N=2$ case the sub-manifolds corresponding to a particular 
level set $v$ are represented by four points, while the sub-manifolds $M_v$ are the fraction of the 
two semi-circles for which $V(q)/N \leq v$. No more topology changes happen in the sub-manifolds 
until the maximum value of the potential $v=0.25$ is reached and the whole circle (sphere) becomes 
accessible. The bottom right panel illustrates the situation for a level $v$ slightly below the 
maximum. At $v=0.25$ a new topology change happens, the two disconnected sectors of the 
sub-manifolds $M_v$ meet each other and complete the manifold, which, for the $N$ particles system, 
is the $(N-1)$-sphere. The maximum of the potential energy per particle, $v_c=1/2N$, tends to zero 
in the thermodynamic limit and coincides with the mean potential energy at the phase transition 
described in (\ref{mean_v}) and figure (\ref{fig.energia.sem.campo}). This shows that the phase 
transition takes place at the level $v_c$ where a topology change in the potential energy 
sub-manifolds happens. This is what is expected according to the topological hypothesis. The 
evolution in the topology of the sub-manifolds $M_v$ as the level $v$ grows also illustrates in a 
nice way how the different sectors of the manifold $M$ become gradually accessible to the physical 
system. From the ground states, the only accessible at zero temperature, two symmetric regions of 
the (hyper)-sphere become gradually accessible in accordance with the symmetry breaking nature of the phase 
transition in this model. In the thermodynamic limit the two regions remain disconnected until the phase transition at $v_c$, 
where the (hyper)-sphere is completed, the two hemispheres connected and the whole configuration 
space manifold becomes accessible to the system. 

This is a simple and completely intuitive example of the topological hypothesis at work. Nevertheless,
 although it was already expected that a topology change must take place in correspondence with a 
phase transition \cite{FrPeSp03}, an yet open question regards the kind of topology change that might 
imply a phase transition. The example of the completely connected spherical model is again useful in 
this respect. At variance with what was observed in previously studied models, in this case it is 
clear that the topology change at the transition is not a strong one, at least as quantified by the change in the Euler characteristic, which is calculated in
the Appendix. There we show that the Euler characteristic is a constant equal to two for 
$1/2N-1/2\leq v < 1/2N$ and jumps to zero at $v_c=1/2N$
when $N$ is even, or does not change at all for $N$ odd. Clearly, from the point of view of the
behavior of the Euler characteristic the change in topology is not a strong one.
In the next section we consider the model in the presence of an external field $H$ which destroys the 
phase transition and analyze the consequences in the topology of the configuration space.

\subsection{Finite external field}
When $H\neq 0$ the energy function in the diagonal basis can be written:
\beq
V = -\frac{J(N-1)}{2N}\, y_1^2 - \sqrt{N} H\, y_1 + \frac{J}{2N}\sum_{i \geq 2} y_i^2
\label{ene_diag_H}
\eeq

The extrema of this function evaluated on the (hyper)-sphere are the same extrema of 
$F = V + \mu(\sum_{i=1}^N y_i^2 - N)$. They are given by the solutions of:
\beqa
\left( 2\mu-\frac{N-1}{N}\right) y_1 & = & H \sqrt{N} \nonumber \\
\left( 2\mu + \frac{1}{N}\right) y_i & = & 0, \,\,\,\,i\neq 1.
\label{critp_F_H}
\eeqa
There are two possibilities as in the zero field case: either $\mu=-1/2N$ or $\mu \neq -1/2N$. 
In the first case the solution is $\{y_1=-H\sqrt{N},\,\sum_{i\geq 2} y_i^2=N(1-H^2)\}$. In the second 
case we obtain $\{y_1=\pm \sqrt{N},\,y_i=0\, \forall i\geq 2\}$. Similarly to the case $H=0$ there are
 two minima and a continuously degenerate maximum. The two minima now correspond to a single absolute 
minimum $\{y_1=\sqrt{N},\,y_i=0\, \forall i\geq 2\}$ and to a local minimum $\{y_1=-\sqrt{N},\,y_i=0\, \forall i\geq 2\}$. 
The corresponding energies per particle are $v_1=-(N-1)/2N-H$ and $v_2=-(N-1)/2N+H$ respectively. 
The other critical point is in fact a critical manifold at the maximum of the energy given by 
$v_3=H^2/2+1/2N$. In the thermodynamic limit the potential energy per particle is $H^2/2$, a level
 that is never reached by the system, as shown in section (\ref{H.dif.0}).

\begin{figure}[ht]
\includegraphics[width=10cm,height=8cm]{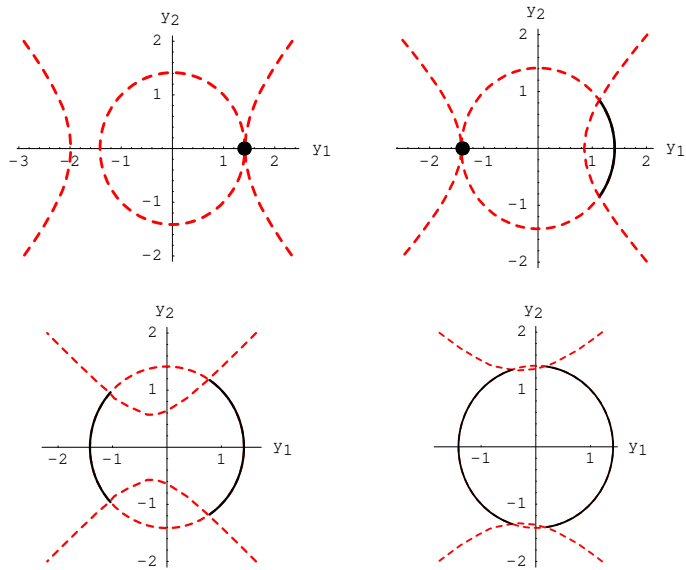}
\caption{(Color online) Evolution of the potential energy manifold of the $N=2$ spherical model for
four levels $v$ and an external field H=0.1. The colored dashed lines represent a particular 
level set of the function $v$ and the spherical constraint (the circle). The real sub-manifolds 
are the continuous black sectors. Top left: the ground state emerges (unique black dot). Top right:
a second local minimum emerges while the accessible sub-manifold is the continuous black arc at the
right of the sphere (circle). Bottom left: at a still higher level $v$ the sub-manifold consists 
of two disconnected arcs. Bottom
right: the highest thermodynamically accessible level $v=0.25$. Note that the sphere is not fully
accessible.}
\label{fig.manifold.H}
\end{figure}

In figure (\ref{fig.manifold.H}) it is shown the evolution of the sub-manifolds $M_v$ for four 
increasing values of $v$. One immediately recognizes the asymmetry introduced by the external field,
 which in these figures is $H=0.1$. The top left panel shows the level where the potential energy 
manifold emerges, corresponding to the ground state, which in this case is unique. At $v=H-1/4$ the 
second minimum touches the sphere. 
This is shown in the top right panel. As the potential energy grows two disconnected regions are 
present, while only one of them is accessible dynamically in the thermodynamic limit. In the bottom 
right panel it is shown the situation at the maximum thermodynamically accessible level. 
The energy per particle of the maximum for $N=2$ is $v_{max}=H^2/2+1/4$ and goes to $H^2/2$ in the 
thermodynamic limit. Nevertheless, the results from the thermodynamics of section (\ref{H.dif.0}) 
predict that the energy per particle reaches a maximum at infinite temperature which is zero, as 
shown in figure (\ref{fig.energia.com.campo}). Consequently the closing of the sphere is never 
reached by the system, the field introduces a gap $\Delta v=H^2/2$ in the energy per particle that 
the system can never cross. The situation at the highest physically accessible level is represented 
in the bottom right panel of figure (\ref{fig.manifold.H}). One is led to the conclusion that the 
only topology changes in the presence of a finite external field are at the levels where the minima 
appear, and that no other topology change takes place at higher levels of $v$, provided one 
restricts the analysis to the physically accessible region of the potential energy manifold. This is
 in agreement with the absence of a phase transition in this case: 
no topology change$\Rightarrow$no phase 
transition. Nevertheless this reading of the results is biased by our {\em a priori} knowledge of
the thermodynamics of the system. In case the thermodynamics would not be known one could be led to 
the wrong conclusion that a phase transition might take place in correspondence with the maximum 
level of the potential energy manifold, where a topology change certainly happens. This suggests 
that topology alone is not enough in order to conclude if a phase transition will or will not take 
place in a particular system.

\section{Conclusions}
The simplicity of the completely connected spherical model allows a critical analysis of some
important open questions regarding the validity of the topological hypothesis. Due to
its high level of symmetry it is possible in this model to intuitively follow the relation
between the topology of the accessible manifold at any given energy level and its physical or
thermodynamic behavior. In particular, the relation between the topology and the symmetry
breaking transition in zero field is nicely illustrated: the phase transition takes place 
in the thermodynamic limit, at the level where the whole manifold, the hyper-sphere, becomes 
accessible. At this level a
simple topology change takes place: the completion of the hyper-sphere. While this is in agreement 
with
a recently proved theorem which asserts the necessity of a topology change in order for a system to
have a phase transition, in this case the change is very small at variance with results from other 
previously studied models. Small topology changes can take place in general with no correlation 
with a phase transition. This is observed for example in the one dimensional XY model 
\cite{CaPeCo03}. 

One can also draw some new conclusions regarding the behavior of the model in the presence of an
external field. In this case the comparison between thermodynamics and topology of the potential
energy manifold can shed some light on the typical behavior of systems in a field. From a
topological point of view little changes occur in the structure and evolution of the sub-manifolds 
$M_v$. For $H\leq J$
the degeneracy between the two minima is broken in a single ground state and a single local
minimum, and a third topology change happens when the hyper-sphere is closed at the highest energy
level. From this behavior one should be tempted to predict a phase transition similar to
that in the zero field case. Nevertheless thermodynamics tells clearly that this is not the case,
there is no phase transition in the presence of a field and the mean potential energy does not 
reach the top level
of the potential energy manifold even at infinite temperature, existing a gap proportional to the
square of the field amplitude. Consequently, to correctly read the information it is
necessary to go through thermodynamics.
It seems unlikely that the knowledge of the topology of the potential energy manifold alone be 
enough in order to predict the existence of a phase transition in a generic many body system. 
The results presented in this work suggest that more information, coming from dynamics, 
is needed. New results from other models are clearly needed in order
to settle this fundamental question \cite{Ka04}. Perhaps the strongest result predicting dynamical
behavior exclusively from a static property is the celebrated Adam-Gibbs relation between
relaxation time scales and configurational entropy in glasses \cite{AdGi65}. It predicts a
divergence of relaxation times when the configurational entropy $S_c$ associated with the number of
minima of the potential energy function goes to zero. Up to our knowledge this prediction has never
been obtained from first principles and even quantitative comparison with experiments and
simulations is not conclusive. 

\appendix*

\section{}
In this Appendix we analyze the topological evolution of the accessible manifold as the potential energy of the system is being increased from its minimum. The complete manifold $M$ for the system is the $(N-1)$-sphere. The function we define on it is $V$, the potential energy function. 
For a given value of the potential energy per particle $v=V/N$, the sub-manifold of accessible configurations is given by $M_v=\{ y \in M| V(y)/N \leq v\}$. We thus analyze the behavior of the Euler characteristic for the sub-manifolds $M_v$ for each $v$, $\chi(M_v)$, which is a topological invariant \cite{NaSe}. It is defined as:
\beq
\chi(M_v)=\sum_{i=0}^{N-1}\, (-1)^i\, b_i,
\eeq
where $b_{i}$ are the Betti numbers or the number of $i$-handles that compose the manifold $M_v$.
According to Morse theory, there is a connection between the topological transitions in a manifold and the critical points of a function defined on it. One of the results of Morse theory concerns $\chi(M_v)$ and is expressed by:
\beq
\chi(M_v) = \sum_{i=0}^{N-1}\, (-1)^i\, b_i = M_{-1}(V).
\eeq
In case $V$ has only isolated critical points $M_{-1}(V)$ is given by:
\beq
M_{-1}(V) =  \sum_{i=0}^{N-1} (-1)^{i}\,m_{i} ,
\eeq
with $m_i$ being the number of critical points of $V$ with index $i$ that belong to $M_v$. The index of a critical point is the number of negative eigenvalues of the Hessian $\mathcal{H}$ of the function at this point.
In order to investigate the critical points of $V(y_1, \ldots ,y_N)$ constrained to the manifold $M$, we make the analysis on the transform $F$ defined in (\ref{function_F}). In what follows we will analyze the cases $H = 0$ and $H\neq 0$. 

\subsection{Zero external field}
As we have seen for $H = 0$ the function $F$ possesses two critical levels.
One of them gives $v=V/N=-(N-1)/2N$ corresponding to two isolated critical points
$\{y_1=\pm \sqrt{N},\,\,y_i=0,\,\,i\neq 1\}$. It is possible to verify that there is no real intersection between the manifold $M$ and the (hyper)-surfaces of constant potential energy for $v<-(N-1)/2N$
 (see top left panel in figure (\ref{fig.manifold})). The accessible sub-manifold $M_v$ for 
$v<-(N-1)/2N$ is empty, and the Euler characteristic is then identically zero: $\chi (v<-(N-1)/2N)=0$
.
The Hessian of $V$ is diagonal in the base $\{ y_i \}$. For the two critical points appearing at 
$v=-(N-1)/2N$ the eigenvalues of $\mathcal{H}$ are given by
\beqa
h_{1} & = & 0 \nonumber\\
h_{i} & = & 1 \ \ \ \ i\neq 1.
\eeqa
The Hessian has no negative eigenvalues. The indexes of both critical points are thus zero. The Euler characteristic is then
$\chi(M_v) = (-1)^{0}\,\, 2 = 2$.
From $v=-(N-1)/2N$ while $v<1/2N$ we have no other critical levels, and thus $\chi$ must remain constant up to $v=1/2N$.

At $v_c=1/2N$ the solutions of equations (\ref{cript_F}) are $\{y_1=0, \sum_{i=2}^{N} y_i^2=N\}$, 
an $(N-2)$-dimensional critical sphere. In fact, since we have already seen that at $v=1/2N$ the manifold completes itself into the (hyper)-sphere $M$, we know that $\chi(M_v)=\chi(M)=\chi(\mathcal{S}^{N-1}), \,\,\, \forall v \geq 1/2N$. The Betti numbers for the sphere are well known: an $N$-sphere is composed of a $0$-handle and of an $N$-handle. The Euler characteristic for the $(N-1)$-sphere is then:
\beq
 \chi(\mathcal{S}^{N-1})= \left \{ \begin{array}{l}
                                                2\,\,\, i\!f\, N\,  odd\\
                                                0\,\,\, i\!f\, N\,  even
                                               \end{array}
                                      \right.
\eeq

Although $\chi(M_v)$ may not change at $v_c$ for odd $N$ one knows that a topology change
takes place on that level. This result is not contradictory since the behavior of a single
topological invariant is not always enough in order to fully characterize the topology of a
manifold.

\subsection{Finite External Field}
For $H > 0$ 
the points $\{y_1=\pm \sqrt{N},\,\,y_i=0,\,\,i\neq 1\}$ still are solutions. However the point
$y_1=+\sqrt{N}$ now corresponds to the level $v_1=-(N-1)/2N-H$ and $ y_1=-\sqrt{N}$ to 
$v_2=-(N-1)/2N+H$.
Since there is no critical value lower than $v_1$, the Euler characteristic for potentials below this level is identically zero: $\chi(M_v)=0, \,\, \forall v < v_1$.
At the critical point corresponding to $v=v_1$ the eigenvalues of the Hessian are:
\beqa
h_{1}&=&0,\nonumber \\
h_{i}&=&1+H, \,\,\,\, i\neq 1.
\eeqa
None of the eigenvalues is negative, thus the critical point is a minimum and has index zero. 
The Euler invariant is then
$\chi(M_v)=(-1)^{0}\,\, 1=1$ for $v_1 \leq v< v_2$. 
The next contribution comes from the critical point $\{ y_1=-\sqrt{N}, y_i=0\,\,\forall i\neq 1\}$ 
at $v=v_2=-(N-1)/2N+H$. At this critical point the eigenvalues of the Hessian are:
\beqa
h_{1}&=&0,\nonumber \\
h_{i}&=&1-H, \,\,\,\, i\neq 1.
\eeqa
We notice that the index of this critical point will depend on $H$. For $H\leq 2$, this critical point has index $0$, hence being a local minimum; however for $H>2$  the index is $N-1$, and the critical point is a maximum. 
At $v=v_2$ the Euler characteristic becomes:
\beq
 \chi(M_v)=\left \{ \begin{array}{ll}
                      2  & \,\,\,\, if\,H\leq 2\\
                      0  & \,\,\,\, if\, H> 2\,\,and\,\,N\,\,even\\
                      2  & \,\,\,\, if\, H> 2\,\,and\,\,N\,\,odd 
                      \end{array}
             \right.\\
\eeq
There is a third solution of the critical point equations which is given by
$\{y_1=-\sqrt{N}H, \sum_{i=2}^{N}y_i=N(1-H^2)\}$. This solution only exists for $H\leq 2$. 
The corresponding critical value of the potential energy is $v_3=H^2/2+1/2N$, 
which is higher than the previous ones for any $H$.
We have thus two different possibilities: 
For $H\leq 2$, $M_v$ coincides with $M$ (the whole (hyper)-sphere) up from $v=v_3$. For $H>2$, $M_v$ coincides with $M$ up from $v=v_2<v_3$. Similarly to what was done in the previous section, we can use this information and simply identify $\chi(M_v)$ with $\chi(M)=\chi(\mathcal{S}^{N-1})$ for either $v \geq v_3$, or $v \geq v_2$.

\begin{acknowledgments}
We would like to thank Lapo Casetti for useful comments and suggestions. 
This work was partly supported by CNPq, Brazil.
\end{acknowledgments}

\end{document}